# ADVANCED ANTENNA TECHNIQUES AND HIGH ORDER SECTORIZATION WITH NOVEL NETWORK TESSELLATION FOR ENHANCING MACRO CELL CAPACITY IN DC-HSDPA NETWORK

Muhammad Usman Sheikh[1], Jukka Lempiainen[1] and Hans Ahnlund[2]

[1]Department of Communications Engineering, Tampere University of Technology
[2]European Communications Engineering Ltd, Tekniikantie 12, Espoo Finland

**ABSTRACT**

*Mobile operators commonly use macro cells with traditional wide beam antennas for wider coverage in the cell, but future capacity demands cannot be achieved by using them only. It is required to achieve maximum practical capacity from macro cells by employing higher order sectorization and by utilizing all possible antenna solutions including smart antennas. This paper presents enhanced tessellation for 6-sector sites and proposes novel layout for 12-sector sites. The main target of this paper is to compare the performance of conventional wide beam antenna, switched beam smart antenna, adaptive beam antenna and different network layouts in terms of offering better received signal quality and user throughput. Splitting macro cell into smaller micro or pico cells can improve the capacity of network, but this paper highlights the importance of higher order sectorization and advance antenna techniques to attain high Signal to Interference plus Noise Ratio (SINR), along with improved network capacity. Monte Carlo simulations at system level were done for Dual Cell High Speed Downlink Packet Access (DC-HSDPA) technology with multiple (five) users per Transmission Time Interval (TTI) at different Intersite Distance (ISD). The obtained results validate and estimate the gain of using smart antennas and higher order sectorization with proposed network layout.*

## 1. INTRODUCTION

Rising trend of packed switched traffic and high capacity requirement in mobile networks have urged the researchers to think about new antenna designs and possible network layouts for future cellular networks. Current and future capacity demands of next generation mobile networks cannot be achieved by using traditional macro cells only. It has been noted several times that macro cells are not able to offer high data rates homogeneously over the entire cell area, and most of the network capacity is lost due to interference coming from the neighbor cells. The increasing demand of new advanced mobile services with different Quality of Service (QoS) requirement in cellular systems has led to the development and evolution of new technologies. Concepts of micro cells and femto cells have been proposed to improve the system capacity in high density traffic areas [2]. However, to reduce the fixed costs such as electricity, transmission, rentals etc., adding new cells and sites should be avoided. Maximum capacity utilization of macro cells should be guaranteed by adopting new network tessellation and by employing possible advanced antenna solutions, including smart antennas. Smart antennas have gained enormous popularity in the last few years, and have been able to grab the attention for its ability to improve the performance of cellular systems [3].

Moreover, cell and system capacities are related to network layout, antennas deployment techniques, orientation and beamwidth of antennas. Directional antennas with optimum electrical or mechanical tilt are used to get required coverage with minimum interference [2]. Antenna configuration i.e. antenna height, azimuth, radiation pattern and beamwidth has deep impact on cell capacity [4– 6]. Different network tessellations have been compared in [7], and it was noted that for 3-sector sites, cloverleaf layout offers the lowest interference level, and thus should have the best cell and system capacity for macro cells. Thus, cloverleaf is a good basis for nominal planning of mobile networks with 3-sector sites. However, for higher order of sectorization, cloverleaf layout cannot be used and new tessellation is needed to combat the problem of interference. Base station antenna configuration needs to be optimized to attain minimum inter cell interference [1], [8]. The conventional cellular concept approach uses fixed beam position with wide beamwidth. Whereas, advanced approach of smart antenna employs multiple narrow beams and beam steering for each user in a cell. Adaptive algorithms form the heart of antenna array processing network. The processor based on different beamforming algorithms does the complex computation for beamforming [9]. Achieved user SINR and user throughput strongly relies on interference management and inter-cell interference avoidance [10]. Handovers between cells due to mobility of user, and software features have their own impact on cell capacity. However, this research work does not deal with these issues.

Over the last decade, services like multimedia messaging, video streaming, video telephony, positioning services and interactive gaming have become an integral part of everyday life. These services are the driving force in reshaping the cellular technologies. Universal Mobile Telecommunication System (UMTS) has been the most popular choice for 3G mobile communication systems, but UMTS had challenges in meeting the requirement of high data rate services. High Speed Downlink Packet Access (HSDPA) was for first time introduced in Release 5 of 3GPP specifications [8], [10]. The evolution of HSDPA continued and later in Release 8, the concept of Dual Cell HSDPA was floated in which the radio resources of two adjacent HSDPA carriers were aggregated with the help of joint scheduler. The main target of DC-HSDPA was not only to improve the user's throughput in the close vicinity of base station rather it also enhances the user's throughput homogeneously over whole cell area. DC-HSDPA offers theoretical peak data rate of 42 Mbps, improved spectral efficiency, and enhanced user experience with low delays or latency [8], [11].

In this paper user's SINR value, average SINR over the cell, mean cell throughput, mean site throughput, user's throughput and user's probability of no data transfer will be taken as merits of performance. Statistical analysis with $10^{th}$, $50^{th}$, $90^{th}$ percentile, and mean value is also presented in this paper. The rest of the paper is organized as follows. Section II deals with theoretical aspects of cell capacity. Section III explains different antenna techniques. Description of simulation tool and environment, simulated cases, and simulation parameters is presented in section IV. Simulation results and their analysis are given in section V. Finally, section six concludes the paper.

## 2. CELLULAR THEORY

### 2.1. Interference and Cell Capacity

Theoretical maximum cell capacity can be estimated by Shannon capacity equation (bits/s) for Additive White Gaussian Noise (AWGN) channel as given in equation (1), [1], [4]

$$C = W \log_2 \left(1 + \frac{S}{N}\right) = W \log_2 \left(1 + {E_b R}/{N_0 W}\right) \quad (1)$$

where $W$ is the bandwidth available for communication, $S$ is the received signal power which can be denoted as energy per information bit $E_b$, multiplied with the information rate $R$. A variable $N$ is the noise power impairing the received signal. The noise power can be defined as noise spectral density $N_0$ multiplied with the transmission bandwidth $W$. Signal to Noise ratio (S/N) can be extended to Signal to Interference plus Noise Ratio (SINR) by including interference from own cell and also co-channel interference coming from neighbor cells. HSDPA is a WCDMA based network, and the total interference is a sum of three interference sources; own (serving) sector signals, other site/sector signals, and thermal noise. In downlink direction, the total interference $I_{DL}$ for any particular user at a given location is given by equation (2), [8]

$$I_{DL} = I_{other} + I_{own} + P_N \quad (2)$$

$$I_{other} = \sum_{i=1}^{k} \frac{Pt_i}{L_i} \quad (3)$$

$$I_{own} = \frac{(Pt_S - S_j)}{L_S} * (1 - \alpha) \quad (4)$$

In equation (2), $P_N$ is a thermal noise power. In equation (3), $I_{other}$ is the total received power from other sectors of the network, and is a sum of other cells interfering sources. $Pt_i$ is a total transmit power and $L_i$ is a path loss for $i^{th}$ neighbor cell. In equation (4), $I_{own}$ is the total received interference from own sector, where $Pt_S$ and $L_S$ are the total transmit power and path loss respectively of serving cell. Where $S_j$ is the received power of HS-PDSCH of the $j^{th}$ user from the serving NodeB. $\alpha$ denotes orthogonality factor. Orthogonality is a measure for level of interference caused by own sector signals. For perfectly orthogonal DL channelization codes $\alpha$ is equals to 1. In HSDPA technology, code orthogonality is partly lost (α < 1) in wireless radio environment due to multipath propagation [8], [12]. The ratio of $I_{other}/I_{own}$ is a commonly used measure of sector overlap and interference in the network layout. The SINR represents the quality of the received signal. In the downlink direction the receiver input, SINR is defined as

$$SINR_{DL} = \frac{S_j}{I_{other} + I_{own} + N} \quad (5)$$

### 2.2. Network layouts and inter cell interference

In initial nominal plan for mobile network, regular network layouts are used for guidance on selection of nominal site location, order of sectorization, and azimuth direction. There is triangular, square, and hexagonal tessellation for 3-sector site, but the most commonly used tessellation is cloverleaf layout as shown in [7]. These tessellations are chose to form continuous coverage of the mobile network. In Fig1a, cloverleaf layout is shown, that is formed by using hexagonal geometry of cell. In cloverleaf layout, all the interfering sites of the first tier of interferer are pointing at the null of serving site. Authors of this paper propose a name "Snow Flake" layout for the enhanced cellular network tessellation for six sector site presented in [13]. Snow flake tessellation is shown in Fig1b. This paper presents a novel network layout for 12-sector site, as shown in Fig.1c and calls it as "Flower" layout. SINR calculations include own cell and neighbor cell interference as given in equation (5). These interferences are related to propagation loss i.e. path loss $L_S$ and $L_i$ between serving NodeB and interfering NodeBs respectively. Especially inter cell interference depends heavily on chosen network layout i.e. how base stations are deployed in a network, antenna configuration, and azimuth. Network layout has significant impact on interference management and hence on capacity of macrocellular network. One way to compare different network layouts or different antenna configurations is to compare the interference coming from neighbor cells to serving cells. It has been shown in [7] that for 3-sector sites, cloverleaf is the most defensive for interference and thus provides high capacity gain. In this article, for 3-sector sites only cloverleaf layout is considered for network simulations.

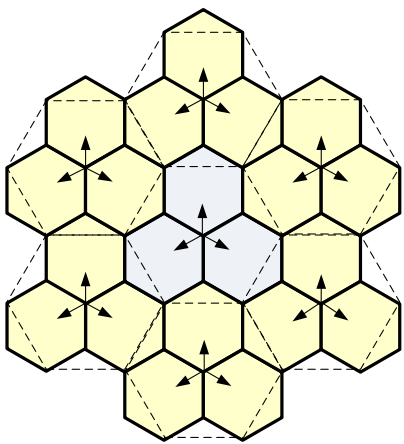

Fig.1. (a) 3-sector "Cloverleaf" layout

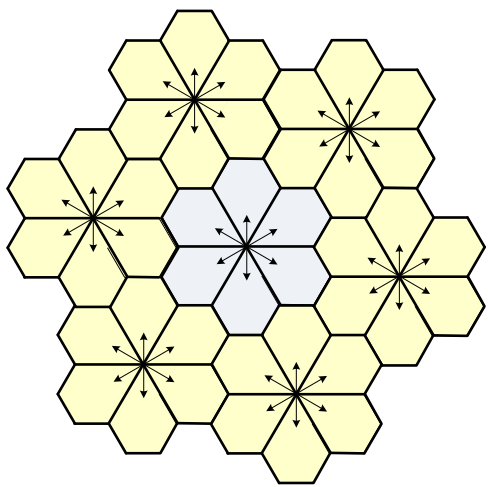

Fig.1. (b) 6-sector "Snow Flake" layout

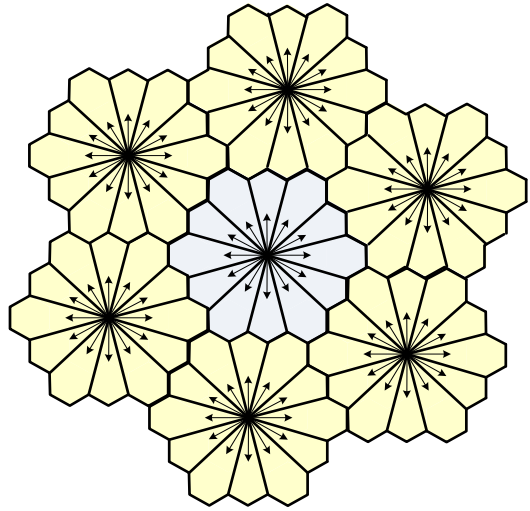

Fig.1. (c) 12-sector "Flower" layout

## 3. ANTENNA THEORY

The functionality of antenna depends on number of factors including physical size of an antenna, impedance (radiation resistance), beam shape, beam width, directivity or gain, polarization etc [14]. By definition, an antenna array consists of more than one antenna element. The radiation pattern of an antenna array depends on number of antenna elements used in array. The more elements there are, the narrower beam can be formed. Planar arrays are capable of making a narrow beam in horizontal as well as in vertical plane. Therefore, planar array beams are also called "Pencil Beams". Smart antennas with ability of beam steering can be constructed by adding "Intelligence" to planar arrays. Smart antenna improves the coverage of cell by concentrating more power in a narrow beam, enhances the cell capacity and offers increased data rates by offering high signal to interference plus noise ratio [15]. By avoiding interference and increasing signal power, smart antenna improves link quality and helps in combating large delay dispersion [16].

In the research work of this paper, three different type of antenna were taken into account i.e.1) Conventional $65^0$ wide beam antenna, 2) Switched beam smart antenna and 3) Full adaptive beam antenna.

### 3.1. Conventional wide beam antenna

In traditional cellular networks, three-sectored approach with $65^0$ wide beam antenna has been in used for long time due to lower interference compared to $120^0$ wide beam antenna. To further improve the performance of fixed wide beam antenna, electrical or mechanical tilting can be used [6]. Base station antennas can be dropped down to building walls but then the propagation environment is not any more related to macro cells, rather shifts to micro cell environment. Other possibility is to modify and narrow the radiation pattern with the help of antenna arrays.

## 3.2. Switched beam smart antenna

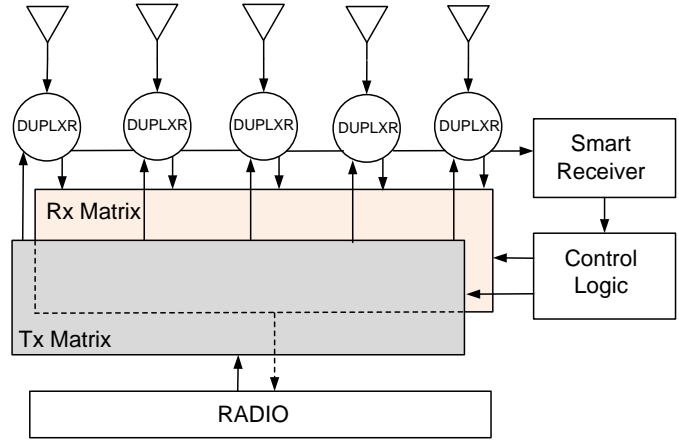

Fig.2. Block diagram of switched beam smart antenna system

Switched beam antenna approach is the extension of conventional cellular sectorisation method, in which single $120^0$ wide macro sector is divided into several micro sectors. A switched beam antenna is a combination of multiple narrow beams in predetermined directions, overlapping over each other. It covers the desired cell area with finite number of narrow fixed beams, where each beam can serve a single user or multiple users [3]. Switched beam antenna does not steer or adapt the beam with respect to the desired signal. In this type of antenna, a RF switch connected to fixed beams controls the beam selection based on the beam-switching algorithm. A switch selects the "Optimum" beam to provide service to mobile station. The optimum beam here refers to the beam that offers the highest SINR value. In some cases, maximum received power for the user can be used as a beam selection criterion. During user mobility, switched beam antenna tracks the user and continuously updates the beam selection to ensure high quality of service [17]. The general block diagram of switched beam smart antenna system is shown in Fig.2 [18].

It consists of an array of antennas that divides the macro sector into several micro sectors. A precise switched beam antenna can be implemented by using "Butler Matrices" [16], [18]. It uses a smart receiver for detecting and monitoring the received signal power for each user at each antenna port. Based on the measurement made by the smart receiver and beam selection algorithm, the control logic block determines the most favorable beam for specific user. The RF switch part governed by the control logic (brain of switched beam antenna) activates the path from the selected antenna port to the radio transceiver. Switched beam antenna offers higher directivity with less interference and thus provides gain over conventional antenna. Theoretically, gain of using switched beam antenna over conventional wide single beam antenna is directly proportional to the number of beams. For a given sector containing U beams, resultant increased gain is given by equation (6) [18]. Switched beam approach is simpler and easier to implement compared to fully adaptive beam approach.

$Gain = 10\text{Log}(U)$     (6)

An example of the horizontal radiation pattern of $65^0$, $32^0$, and $16^0$ HPBW antenna is depicted in Fig.3a, 3b and 3c respectively. Radiation pattern of seven switched beam antenna with each beam of $8^0$ HPBW is shown in Fig.3d.

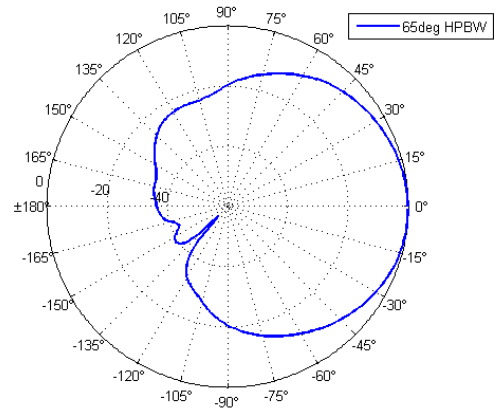

Fig.3. (a) Radiation pattern of conventional $65^0$ beamwidth antenna used in 3-sectored site

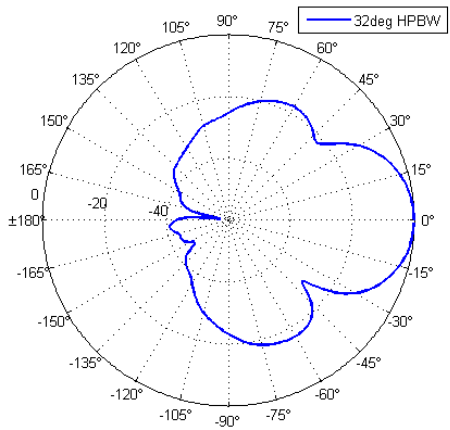

Fig.3. (b) Radiation pattern of narrow $32^0$ beamwidth antenna used in 6-sectored site

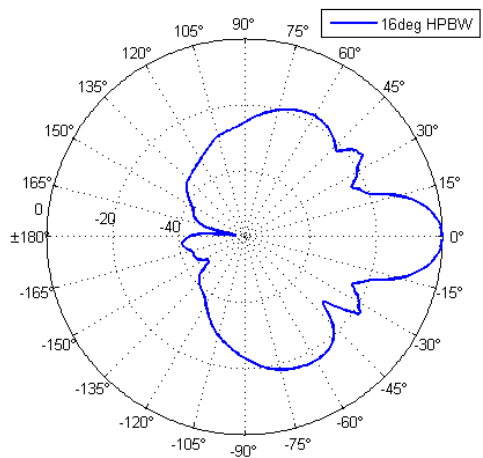

Fig.3. (c) Radiation pattern of narrow $16^0$ beamwidth antenna used in 12-sectored site

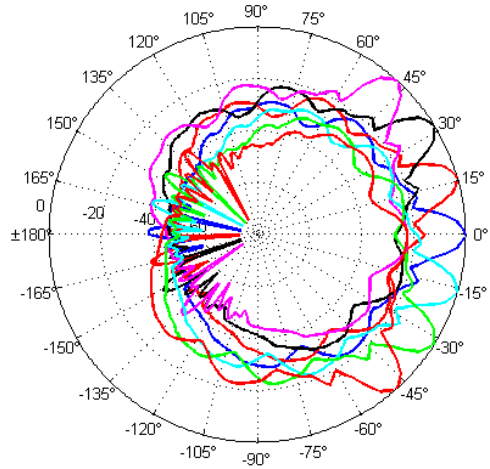

Fig.3. (d) Radiation pattern of switched beam antenna with seven beams of $8^0$ HPBW

### 3.3. Full adaptive beam antenna

Adaptive antenna exploits the array of antenna elements to achieve maximum gain in desired direction while rejecting interference coming from other directions. Adaptive antennas are more complex than multi beam switched antennas. While butler matrices are operating on the RF domain, adaptive antennas use a linear combination of signals, and process them in the baseband. Adaptive antenna can steer its maxima and nulls of the array pattern in nearly any direction in response to the changing environment [16]. The basic idea behind adaptive antenna is the same as in switched beam antenna i.e. to maximize the SINR values. While the multiple switched beam antennas have a limited selection of directions to choose the best beam, an adaptive antenna can freely steer its beam in correspondence to the location of user. Smart antenna employs Direction of Arrival (DOA) algorithm to track the signal received from the user, and places nulls in the direction of interfering users and maxima in the direction of desired user [19]. On the other hand, since adaptive antennas needs more signal processing, multiple switched beam antennas are easier to implement and have the advantage of being simpler, and less expensive compared to adaptive antennas. The overall capacity gain of smart antennas is expected to be in the range of 100% to 200%, when compared with conventional antennas [3].

Beam forming algorithms used in adaptive antennas are generally divided into two classes with respect to the usage of training signal i) Blind Adaptive algorithm and ii) Non-Blind Adaptive algorithm [20]. In a non-blind adaptive beam forming algorithm, a known training signal d(t) is sent from transmitter to receiver during the training period. The beamformer uses the information of the training signal to update its complex weight factor. Blind algorithms do not require any reference signal to update its weight vector; rather it uses some of the known properties of desired signal to manipulate the weight vector. Fig.4 shows the generic beam forming system based on non-blind adaptive algorithm, which requires a training (reference) signal [19].

The output of the beamformer at time $n$, $y(n)$, is given by a linear combination of the data at the $k$ antenna elements. The baseband received signal at each antenna element is multiplied with the weighting factor which adjusts the phase and amplitude of the incoming signal accordingly. The sum of this weighted signal results in the array output $y(n)$. On the basis of adaptive algorithms, entries of weight vector $\mathbf{w}$ are adjusted to minimize the error $e(n)$ between the training signal $d(n)$ and the array output $y(n)$. The output of the beamformer $y(n)$ can be expressed as given in equation (7), [20]

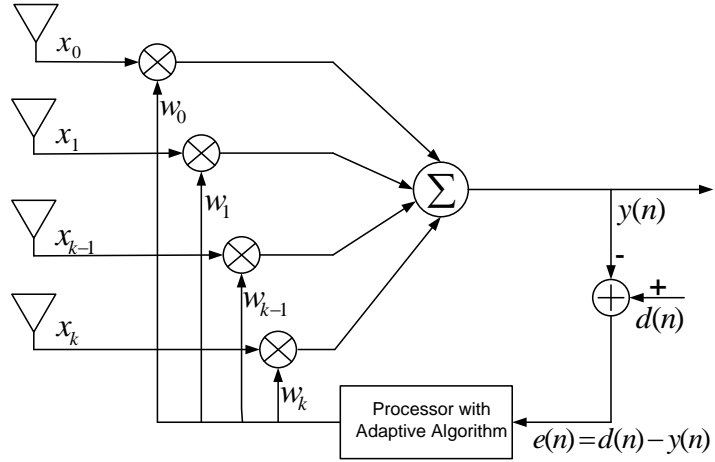

Fig.4. Block diagram of adaptive beamforming system

$y(n) = \mathbf{w}^H(n)\mathbf{x}(n)$ (7)

$\mathbf{w}(n) = [w_1(n) \ w_2(n) \ \ldots \ldots w_{k-1}(n) \ w_k(n)]$ (8)

$\mathbf{x}(n) = [x_1(n) \ x_2(n) \ \ldots \ldots x_{k-1}(n) \ x_k(n)]$ (9)

$e(n) = d(n) - y(n)$ (10)

where $\mathbf{w}(n)$ is the weight vector with $w_k(n)$ a complex weight for $k$th antenna element at time instant $n$, and $[.]^H$ denotes Hermitian (complex conjugate) transpose. $x_k(n)$ is the received baseband signal at $k$th antenna element [9], [20]. Least Mean Square (LMS), Normalized Least Mean Square (NLMS), Recursive Least Squares (RLS), and Direct Matrix Inversion (DMI) are examples of non-blind adaptive algorithm, whereas Constant Modulus Algorithm (CMA) and Decision Directed (DD) algorithms are examples of blind adaptive algorithm [9], [19-20]. These beamforming algorithms have their own pros and cons as far as their computational complexity, convergence speed, stability, robustness against implementation errors and other aspects are concerned.

## 4. SYSTEM SIMULATIONS

### 4.1. Simulation Environment

MATLAB was used as a simulation tool for carrying a campaign of simulations. Monte Carlo type of simulation was done with 5000 iterations with multiple users. It was aimed to model a network as realistic as possible. All system simulations for three sectored sites were done with macro cell cloverleaf layout. Snow flake and flower tessellation was selected for 6-sector 12-sector sites respectively. Base station grid of 19 sites was built, where single middle site in the middle has six sites in the first tier of interferer, and 12 sites in the second tier of interferer. All the interfering sites were at equal Intersite Distance (ISD) as shown in Fig.5(a,b,c), with same site parameters. Base station antenna height was set to 25m, which is typical value in city centre areas where 5-7 floor buildings exists. Power required for common pilot channel and signaling was taken into account. Frequency band of 2100MHz was used in simulations because DC-HSDPA system was selected as an example technology. Simulations were done with flat terrain, and Okumura-Hata model was used for calculating path loss between user and NodeB. Fading component is modelled with log normal distribution having zero mean and 5dB of standard deviation. Orthogonality factor used in equation (4) for computing own cell interference follows Gaussian curve with maximum of 0.97 at site location and 0.7 at cell edge.

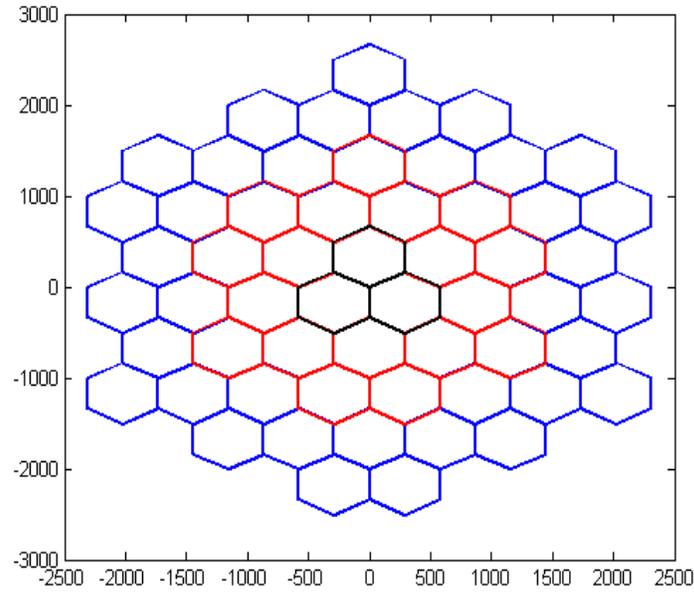

Fig.5. (a) Grid of nineteen 3-sector sites used in simulation with clove-leaf topology

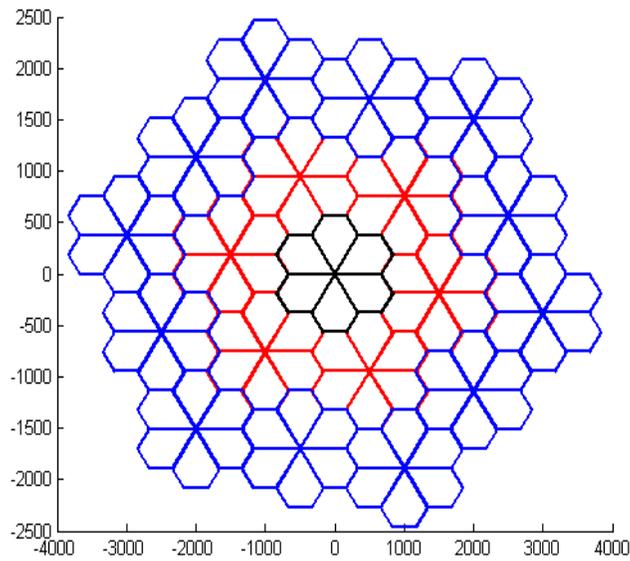

Fig.5. (b) Grid of nineteen 6-sector sites used in simulation with snow flake topology

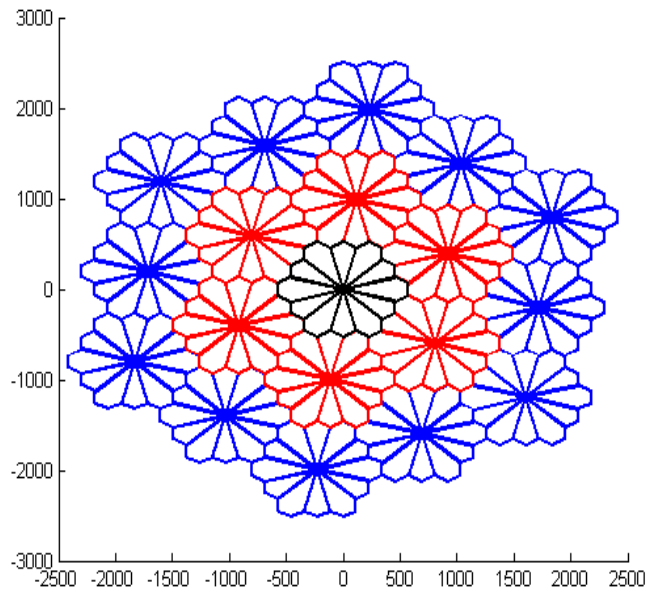

Fig.5. (c) Grid of nineteen 12-sector sites used in simulation with flower topology

## 4.2. Simulation cases and simulation procedure

Following three cases were considered for simulations.

- **3 Sector:** It is the most common scenario in which each site has three sectors and every sector has single $65^0$ half power beamwidth antenna. This acts a reference case for comparing with higher order sectorization and advanced antenna case. Fig.3a shows the radiation pattern of an antenna used for simulations, with no electrical or mechanical tilt, and with maximum antenna gain of 15.39dB.
- **6 Sector:** It is the case in which each site has six sectors, and every sector has single $32^0$ half power beamwidth antenna. Fig.3b shows the radiation pattern of an antenna used for simulations, with no electrical or mechanical tilt, and with maximum antenna gain of 18.20dB.
- **12 Sector:** In this case, each site comprises of 12 narrow sectors, and every sector has $16^0$ HPBW antenna. Fig.3c shows the radiation pattern of an antenna used for simulations, with no electrical or mechanical tilt, and with maximum antenna gain of 21.15dB.
- **7 Switched beams:** This case represents multiple fixed switched beam scenario, where single sector is covered by seven potential narrow beams. Each narrow beam has eight degree HPBW with a spacing of $16^0$ between the beams as shown in Fig.3d. Only that beam which has smallest deviation angle with respect to its main beam for user becomes active for that particular user. No down tilting was assumed, and each beam has maximum antenna gain of 23.55dB.
- **Adaptive beam:** In this last scenario, adaptive antennas are used to form an accurate beam for each individual user. In this scenario, narrow beam of six degree in the horizontal plane is steered precisely to the serving user, keeping user in the middle of the beam for maximum gain. Adaptive antenna have maximum gain of 24.5dB.

Key parameters related to DC-HSDPA systems used in simulations are presented in Table I. For each iteration, 5 users with full traffic buffer in each cell were created. Users were homogenously spread over the whole cell area on the flat terrain. In this simulation, DC-HSDPA serves five code multiplexed users per Transmission Time Interval (TTI). Out of total 16 codes, maximum of 15 codes were available for High Speed Physical Downlink Shared Channel (HS-PDSCH). Total transmission power for HS-PDSCH and available codes were equally distributed among the five users in each TTI. In the serving cell to compute the received signal value, Okumura-Hata model was used to calculate the path loss between the user and serving NodeB. Simulator supports Adaptive Modulation and Coding (AMC), and in these simulations eight different Modulation and Coding Schemes (MCS) were considered with 64QAM 5/6 coding rate as highest and QPSK 1/2 coding rate as lowest possible MCS. As throughput is the function of SINR, hence later SINR information was employed to compute each user throughput. Cell throughput in each TTI is the sum of individual users' throughput. Post processing of data was done to get the results in refined form.

Table I. General DC-HSDPA simulation parameters

| *Parameter* | *Unit* | *Value* |
|---|---|---|
| **DC-HSDPA Downlink** | | |
| Users per TTI | No. | 5 |
| Operating frequency band | MHz | 2100 |
| Bandwidth | MHz | 5 + 5 |
| Chip rate | Mcps | 3.84 |
| Total HS-PDSCH Codes | No. | 15 |
| Max HS-PDSCH power | dBm | 41.63 |
| HS-SCCH power | dBm | 26 |
| Processing gain | dB | 12.04 |
| HSDPA loading | % | 70 |
| Interference margin | dB | 5.2 |
| UE noise figure | dB | 8.0 |
| Downlink activity factor | | 1.0 |

## 5. SIMULATION RESULTS AND ANALYSIS

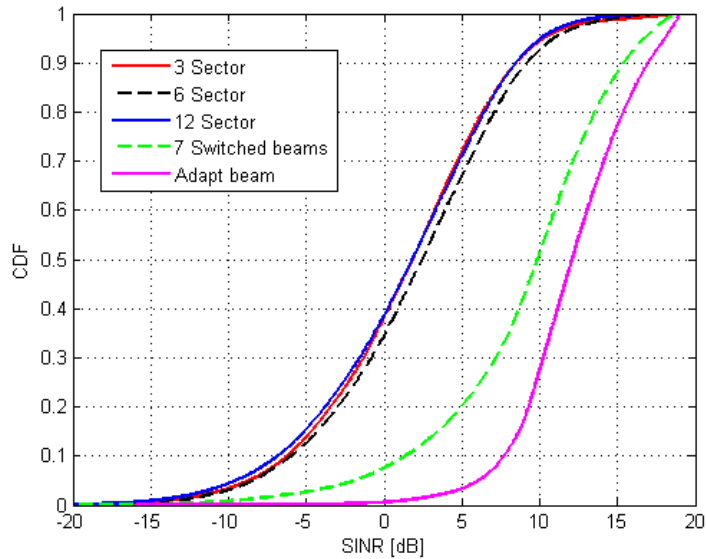

Fig.6. CDF plot of user SINR with 5 users per TTI at 1000m ISD

Fig.6 shows the Cumulative Distribution Function (CDF) of the user SINR with 5 users per TTI at 1000m ISD for different cases. Clearly switched beam antenna shows better performance in terms of offering higher SINR compared to $65^0$, $32^0$, and $16^0$ wide beam antenna used in 3-sector, 6-sector and 12 sector sites respectively. But adaptive beam antenna outperforms and shows superior performance compared to all other cases. By analyzing the curves shown in Fig.6 it can be deduced that adaptive and switched beam antennas served the purpose of improving user experience by reducing the interference and enhancing the received SINR. The CDF curve of SINR for the case of adaptive beam is on the extreme right position, indicating that the SINR for the users is improved on average. It is also important to note that the average user SINR does not deteriorate by increasing the order of sectorization and almost similar performance is shown by 3-sector, 6 sector and 12-sector sites. However, 6-sector site offers slightly better performance compared to 3 and 12-sector sites. Adaptive beam antenna performed well in the close vicinity of the NodeB as well as near the cell edge, as 80% of the samples are concentrated in a narrow range of 9.12dB, starting from 7.72dB to 16.84dB of user SINR. But for other traditional antenna cases, eighty percent of SINR values has wide span and spread over the range of around 14.96dB, starting from -6.3 to 8.66dB. Statistical analysis of user SINR is presented in Table II.

Table II. Statistical Analysis of User SINR

|  | *10 percentile user SINR (dB)* | *50 percentile user SINR (dB)* | *90 percentile user SINR (dB)* | *Mean user SINR (dB)* | *STD user SINR (dB)* | *Relative SINR gain (dB* |
|---|---|---|---|---|---|---|
| 3-Sector | -6.22 | 1.83 | 8.51 | 1.44 | 5.89 | 0 |
| 6-Sector | -5.99 | 2.44 | 9.22 | 1.98 | 5.96 | 0.54 |
| 12-Sector | -6.87 | 1.78 | 8.50 | 1.23 | 6.05 | -0.21 |
| 7 Switched beam | 1.36 | 9.83 | 15.41 | 8.94 | 5.72 | 7.50 |
| Adaptive beam | 7.72 | 12.10 | 16.97 | 12.07 | 3.73 | 10.63 |

Relative SINR gain shown in Table II is the relative gain in dB with respect to the mean SINR value of 3-sector case. It was learned that adaptive and switched beam antennas offer 10.63dB and 7.50dB respectively better user SINR compared to traditional wide antenna used in 3-sector site at 1000m intersite distance.

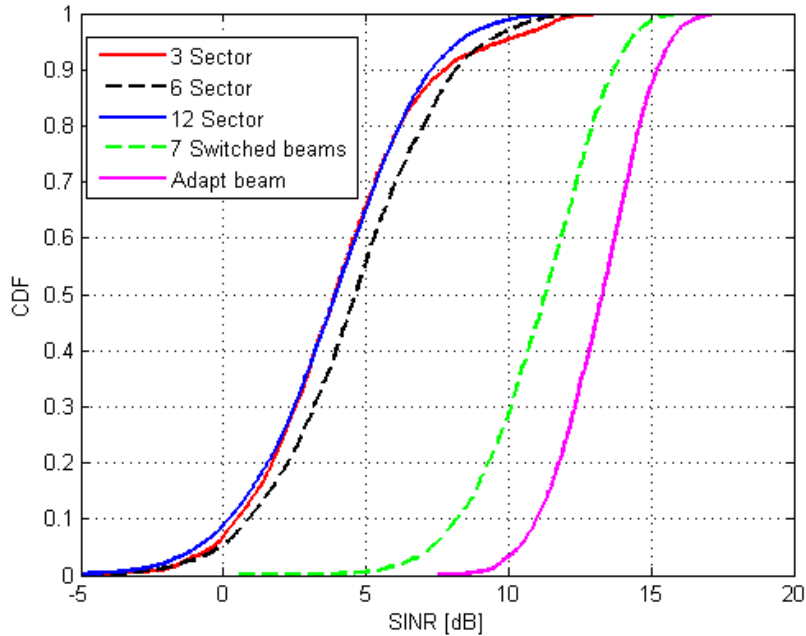

Fig.7. CDF plot of cell SINR with five users per TTI at 1000m ISD

Fig.7 shows the cumulative distribution function of SINR value averaged over the whole cell with 5 users per TTI at 1000m ISD for different simulated cases. Averaged SINR value over the whole cell area in each iteration of Monte Carlo simulation was obtained by adding the linear SINR value of each user and then divide the sum by number of users served per TTI. It can be seen that 6-sector deployment helps in improving the cell SINR by a small margin of 0.51dB only compared to 3-sector deployment, but a significant improvement of 7.02dB and 9.11dB is witnessed in case of switched beam and adaptive beam case respectively. Smart antennas not only improve the user experience rather they improve the overall cell SINR as well. It is also evident that the multiple switched beam antenna offers improvement in SINR but the difference is smaller compared to adaptive antenna. Statistical analysis of cell SINR is given in Table III.

Table III. Statistical Analysis of SINR over whole Cell

|  | *10 percentile cell SINR (dB)* | *50 percentile cell SINR (dB)* | *90 percentile cell SINR (dB)* | *Mean cell SINR (dB)* | *STD cell SINR (dB)* | *Relative SINR gain (dB* |
|---|---|---|---|---|---|---|
| 3-Sector | 0.49 | 3.91 | 7.79 | 4.07 | 2.94 | **0** |
| 6-Sector | 0.91 | 4.63 | 8.14 | 4.58 | 2.85 | **0.51** |
| 12-Sector | 0.21 | 3.94 | 7.31 | 3.84 | 2.80 | **-0.23** |
| 7 Switched beam | 8.19 | 11.28 | 13.68 | 11.09 | 2.14 | **7.02** |
| Adaptive beam | 10.96 | 13.29 | 15.21 | 13.18 | 1.63 | **9.11** |

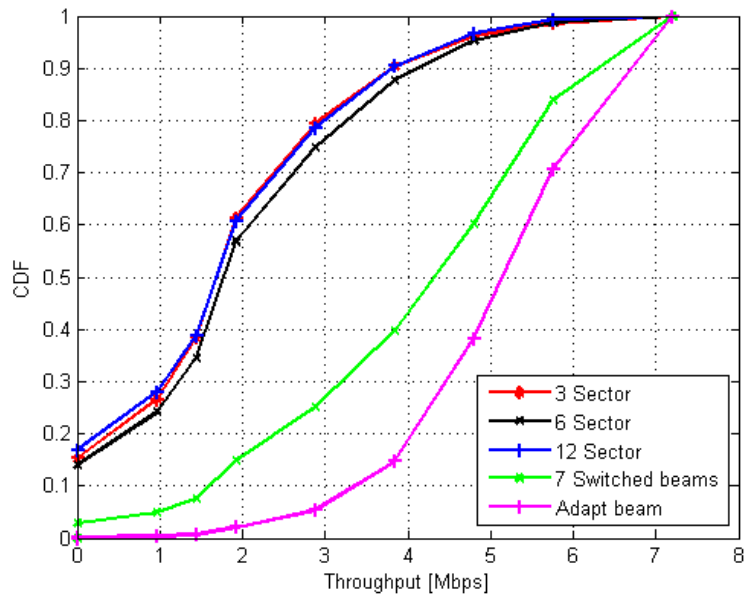

Fig.8. CDF plot of user throughput with 5 users per TTI at 1000m ISD

Fig.8. shows the CDF of the user throughput of DC-HSDPA network with 5 users per TTI at 1000m ISD for different antenna solutions. Eight marks on CDF plots represent eight different MCS. As equal codes and equal power was distributed among the users, therefore high throughput samples show that high modulation and coding scheme was used by the user. High MCS are less robust against interference and thus have high requirement of SINR. It is interesting to note that around 4.5% of the users were able to adapt 64QAM in 3-, 6-, and 12-sector case, whereas this number was raised to 39.98% and 61.8% by switched and adaptive beam antennas respectively. As seen from the results, more than 85% of the samples with adaptive beam were obtained with three highest MCS. Samples of zero throughputs in CDF plots represent the users with no data transfer due to very low SINR. It was also noted that single wide beam antenna keeps the probability of no data transfer at almost 15%. Whereas, switched beam antenna and adaptive beam antenna show remarkable improvement in probability of no data transfer and kept it at negligible level of 2.88% and 0.16% respectively. These results clearly indicate the impact of advanced antenna techniques in improving the user experience, when other cells are heavily loaded and are severely interfering the serving cell.

Fig.9 shows the CDF of cell throughput achieved by using DC-HSDPA with equal power and equal codes allocation for different network tessellation and antenna techniques. Cell throughput in each TTI was computed by summing the individual throughput of the served users. Like in previous results, case adaptive beam lead the comparison and shows extra ordinary performance compared to other network tessellations and antenna types in terms offering higher cell throughput. Almost identical cell throughput is achieved in 3-sector and 12-sector case, but 6-sector offers slightly better capacity. High SINR values showed in Fig.7 is translated into high throughput values in Fig.9. Adaptive beam antenna exhibits better performance and offers 27.99Mbps of average cell throughput compared to 22.81Mbps by switched beam case. 10 percentile cell throughput shows that 90% of the cell throughput samples with adaptive beam were above 24Mbps, and with switched beam 90% of the samples were above 17.28Mbps. Relative throughput gain is the relative gain in percentage value compared to 3-sector case. In [9], it was expected to get 100-200% improvement in cell capacity by smart adaptive antennas, and the results shown in Fig.9 are in line with the expectation. Adaptive beam shows a significant relative gain of 156.7%, however switched beam have relative gain of 109.27%. Statistical analysis of cell throughput is shown in Table IV.

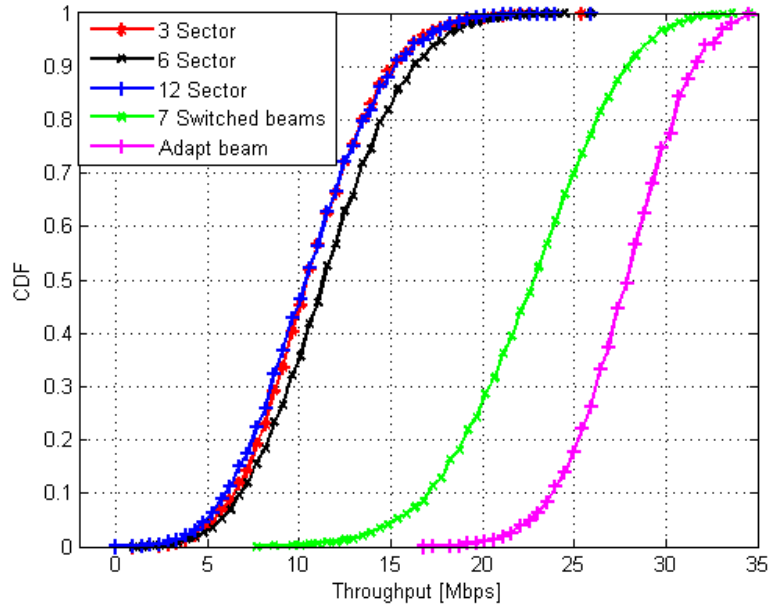

Fig.9. CDF plot of cell throughput with five users per TTI at 1000m ISD

Table IV. Statistical Analysis of Cell Throughput

|  | *10 percentile cell throughput (Mbps)* | *50 percentile cell throughput (Mbps)* | *90 percentile cell throughput (Mbps)* | *Mean cell throughput (Mbps)* | *STD cell throughput (Mbps)* | *Relative throughput gain (%)* |
|---|---|---|---|---|---|---|
| 3-Sector | 6.72 | 10.56 | 15.36 | 10.90 | 3.40 | **0** |
| 6-Sector | 7.20 | 11.52 | 16.32 | 11.72 | 3.67 | **7.52** |
| 12-Sector | 6.24 | 10.56 | 15.36 | 10.77 | 3.59 | **-1.20** |
| 7 Switched beams | 17.28 | 23.04 | 28.32 | 22.81 | 4.23 | **109.27** |
| Adaptive beam | 24.0 | 28.32 | 31.68 | 27.99 | 3.03 | **156.70** |

Fig.10 shows the mean cell throughput of the DC-HSDPA cell with five users per TTI against the intersite distance for different cases. The trend of the sectored antenna cases and switched beam antenna case show that average cell throughput increases by increasing the intersite distance. Small intersite distance corresponds to small cells; hence, the high interference coming from the neighbor cells limit the cell throughput. The variations in the cell throughput for all cases except the adaptive antenna case were caused by the fact that larger the intersite distance, smaller will be the impact of interfering cells and hence larger will be the achieved average cell throughput. However, for adaptive antenna case cell throughput is inversely proportional to the intersite distance. The results show that a deployment of smart antennas significantly enhances the average cell throughput irrespective of the ISD. The highest cell throughput was achieved with adaptive beam antenna at small ISD of 250m. However, the worst capacity is offered by 12-sector antenna at 3000m ISD. It means higher order of sectorization not necessarily offers better cell throughput.

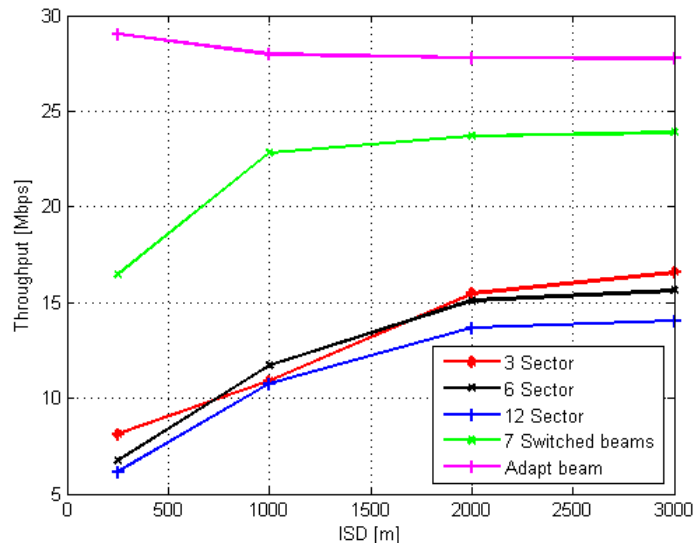

Fig.10. Mean cell throughput with five users per TTI against ISD

Fig.11 shows the achieved mean site throughput for DC-HSDPA system against the intersite distance for different cases. As seen in Fig.11, applying higher order sectorization and deploying advance antenna techniques provides significant throughput gain over traditional 3-sector site topology. Relative site throughput gain for 6-sector and 12-sector topology is higher at large intersite distances than small ISD. With respect to the reference case of 3-sector site at 1000m ISD, when intersite distance is reduced to 250m (small cell) for 3-sector site, mean site throughput is reduced by 25.41%. However, a relative throughput gain of approximately 23.67% and 125.69% is achieved by 6-sector and 12-sector sites respectively at 250m ISD, which is comparatively small compared to 164.04% and 401.65% by 6 and 12-sector sites respectively at 2000m ISD. Adaptive antenna beam outperformed at 250m ISD and was found more effective at small ISD. More detailed analysis of site throughput and the relative gain is presented in Table V. Relative gains shown in Table V are with respect to reference case 3-sector site at 1000m ISD. Negative value of gains means inferior performance.

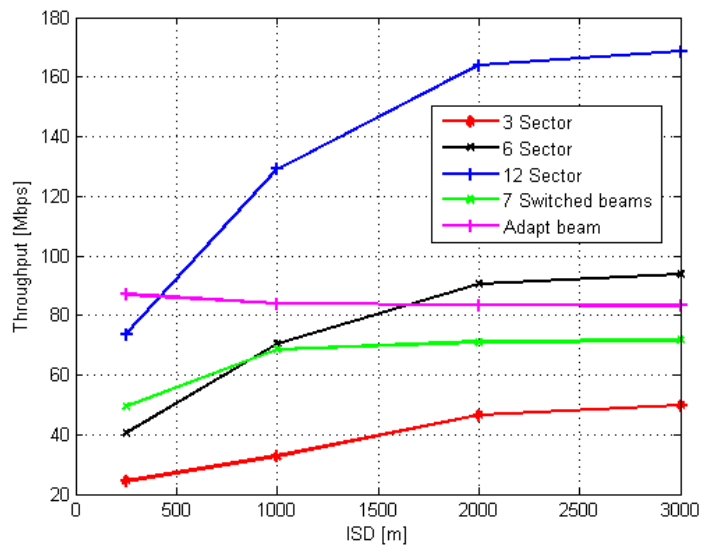

Fig.11. Mean site throughput with five users per TTI against ISD

Table V presents the average downlink throughput and relative sector (cell) gain with respect to 3-sector at 1000m ISD (reference case).

Table V. Statistical Analysis of Cell Throughput

|  | Mean cell throughput (Mbps) | Relative cell throughput gain (%) | Mean site throughput (Mbps) | Relative site throughput gain (%) |
|---|---|---|---|---|
| **ISD = 250 meter** | | | | |
| 3-Sector | 8.13 | **-25.41** | 24.39 | **-25.41** |
| 7 Switched beams | 16.48 | **51.20** | 49.44 | **51.20** |
| Adaptive beam | 29.01 | **166.15** | 87.03 | **166.15** |
| 6-Sector | 6.74 | **-38.17** | 40.44 | **23.67** |
| 12-Sector | 6.15 | **-43.57** | 73.80 | **125.69** |
| **ISD = 1000 meter as reference** | | | | |
| 3-Sector | 10.90 | **0** | 32.70 | **0** |
| 7 Switched beams | 22.81 | **109.27** | 68.43 | **109.27** |
| Adaptive beam | 27.98 | **156.70** | 83.94 | **156.70** |
| 6-Sector | 11.72 | **7.52** | 70.38 | **115.23** |
| 12-Sector | 10.77 | **-1.20** | 129.24 | **295.23** |
| **ISD = 2000 meter** | | | | |
| 3-Sector | 15.49 | **42.11** | 46.47 | **42.11** |
| 7 Switched beams | 23.71 | **117.53** | 71.13 | **117.53** |
| Adaptive beam | 27.80 | **155.05** | 83.40 | **155.05** |
| 6-Sector | 15.10 | **38.53** | 90.60 | **177.06** |
| 12-Sector | 13.67 | **25.41** | 164.04 | **401.65** |
| **ISD = 3000 meter** | | | | |
| 3-Sector | 16.58 | **52.11** | 49.74 | **52.11** |
| 7 Switched beams | 23.89 | **119.17** | 71.67 | **119.17** |
| Adaptive beam | 27.75 | **154.58** | 83.25 | **154.58** |
| 6-Sector | 15.63 | **43.39** | 93.78 | **186.79** |
| 12-Sector | 14.06 | **28.99** | 168.72 | **415.97** |

## 6. CONCLUSION

In this article, we investigated advance antenna techniques along with different network tessellations including cloverleaf topology for 3-sector sites, snow flake topology for 6-sector sites and proposed a novel flower topology for 12-sector sites in DC-HSDPA network. Impact of intersite distance on the performance of higher order sectorization and on the performance of adaptive and switched beam antenna was also taken into account. A comprehensive set of simulation results were demonstrated together with a performance analysis. Post simulation analysis confirms that the capacity gain achieved by higher order sectorization and switched beam antenna increases by increasing the ISD. However, adaptive beam antenna also significantly improves the cell SINR and cell throughput, but adaptive antenna is more effective in small cells compared to large ISD. The simulation results revealed that the average cell SINR does not deteriorate much by having higher order sectorization, however 6-sector site provides around 0.5dB better cell SINR compared to 3-sector site. At 1000m ISD, the cell SINR is improved by approximately 7.02dB and 9.11dB when switched beam and adaptive beam antennas were used respectively compared to traditional 3-sector site with $65^0$ beamwidth antenna. Significant improvement was also witnessed in terms of average cell throughput, it was found that the average cell throughput increased by 109.3% with multiple switched beam antenna, and 156.7% with adaptive beam antenna compared to 3-sector site at 1000m ISD. Adaptive beam antenna outperformed and offered high SINR, high throughput with low probability of no data transfer. Multiple switched beam antenna showed better performance compared to single beam antenna but inferior to adaptive beam. Three-sector and higher order sectorization offer almost 15% of probability of no data transfer for the user at 1000m ISD. Switched beam antenna helps in improving the probability of no data transfer and kept it at almost 2.88%, but adaptive antenna significantly improved probability of no data transfer and brought it down to 0.16% at 1000m ISD. Simulation results revealed that the user experience and the macro cell capacity can be significantly improved by deploying smart antennas. Higher order sectorization does not improve much the cell (sector) capacity, but definitely offers higher site capacity. Especially at large ISD, high order sectorization is more effective and significantly increases the site capacity. To avoid the deployment of small cells, usage of adaptive and switched beam antennas or higher order sectorization can be considered as an alternate choice.

The results were obtained by using semi-statistic simulations with Okumura-Hata propagation model, and thus may cause offset type of error in all results. However, the obtained results are still comparable with each other to show capacity improvements. For future work, it would be interesting to see the performance of fixed switched beam antenna with narrower and even more number of beams in a cell, as in this paper seven beams of $8^0$ were considered in each cell.

## ACKNOWLEDGEMENTS

Authors would like to thank Tampere University of Technology and European Communications Engineering (ECE) Ltd. and Tampere Doctoral Program of Information Science and Engineering (TISE) for supporting the research work of this paper.

**Authors**

**Muhammad Usman Sheikh** was born in Rawalpindi, Pakistan, in 1983. He received the B.S. degree in Electrical Engineering (Telecommunication) from COMSATS Institute of Information Technology, Islamabad, Pakistan, in 2006. He received the M.S. degree in Radio Frequency Electronics from Tampere University of Technology (TUT), Finland, in 2009. Currently, he is working towards the Dr. Tech. degree. His main research interests are radio network planning aspects of GSM/UMTS/LTE networks and traffic handling in the multilayer networks. He is student member of IEEE.

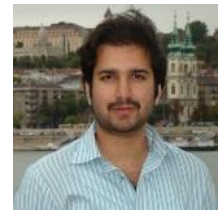

**Jukka Lempiäinen** was born in Helsinki, Finland, in 1968. He received an MSc, Lic Tech, Dr Tech. all in Electrical Engineering, from Helsinki University of Technology, Espoo, Finland, in 1993, 1998 and 1999, respectively. He is a Senior Partner and the President of European Communications Engineering Ltd. He has altogether more than ten years experience in GSM based mobile network planning and consulting. Currently, he is also a part-time Professor of the Telecommunications (Radio Network Planning) at Tampere University of Technology, Finland. He has written two international books about GSM/GPRS/UMTS cellular radio planning, several international journal and conference papers and he has three patents. He is URSI National Board Member, Finland.

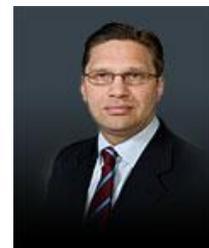

**Hans Ahnlund** was born in Porjus, Sweden in 1968. He received his M.Sc. (EE) degree from Lund University of Technology in 1994. He is Vice President and board member of European Communications Engineering. He is one of the authors of the book 'UMTS Radio Network Planning, Optimization and QoS Management'. He holds two patents in the wireless technology sector.

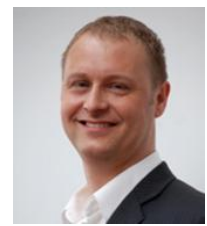